\documentclass[twocolumn,showpacs,preprintnumbers,amsmath,amssymb]{revtex4}

\usepackage{graphicx}
\usepackage{dcolumn}
\usepackage{bm}
\begin{document}

\preprint{}

\title{Experimental quantum key distribution without monitoring signal disturbance}

\author{Hiroki Takesue$^{1}$} \email{takesue.hiroki@lab.ntt.co.jp}
\author{Toshihiko Sasaki$^{2}$} \email{sasaki@qi.t.u-tokyo.ac.jp}
\author{Kiyoshi Tamaki$^{1}$} 
\author{Masato Koashi$^{2}$}
\affiliation{%
$^1$NTT Basic Research Laboratories, NTT Corporation\\
3-1 Morinosato Wakamiya, Atsugi, Kanagawa, 243-0198, Japan\\
$^2$Photon Science Center, Graduate School of Engineering, The University of Tokyo\\
Bunkyo-ku, Tokyo 113-8656, Japan\\
}%




\maketitle



{\bf 
Since the invention of Bennett-Brassard 1984 (BB84) protocol \cite{bb84}, many quantum key distribution (QKD) protocols have been proposed \cite{e91,b92,huttner,cv,kyo,sarg,cow} and some protocols are operated even in field environments \cite{secoqc,tokyoqkd}. 
One of the striking features of QKD is that QKD protocols are provably secure unlike cryptography based on computational complexity assumptions \cite{rmp}. It has been believed that, to guarantee the security of QKD, Alice and Bob have to monitor the statistics of the measurement outcomes which are used to determine the amount of the privacy amplification to generate a key \cite{rmp2}. Recently a new type of QKD protocol, called round robin differential phase shift (RRDPS) protocol, was proposed, and remarkably this protocol can generate a key {\it without} monitoring any statistics of the measurement outcomes \cite{rrdps}. Here we report an experimental realization of the RRDPS protocol. We used a setup in which Bob randomly chooses one from four interferometers with different pulse delays so that he could implement phase difference measurements for all possible combinations with five-pulse time-bin states. 
Using the setup, we successfully distributed keys over 30 km of fiber, making this the first QKD experiment that does not rely on signal disturbance monitoring. 
}

One may wonder what makes such a fundamentally big difference in the RRDPS protocol. To answer this question, let us briefly explain the crux of the RRDPS protocol. The RRDPS protocol is similar to the original differential phase shift (DPS) protocol \cite{kyo} in the sense that both protocols employ trains of $L$ coherent pulses as an information carrier sent by the sender (Alice), and each of the pulses is phase modulated to incur either 0 or $\pi$ optical phase shift according to Alice's random choice. A crucial difference of the RRDPS protocol from the DPS protocol lies in the receiver (Bob)'s detection unit. In the DPS protocol, Bob's setup reads out the relative phase of two adjacent pulses, while that of the RRDPS protocol randomly picks up two pulses among $L$ pulses and reads out the relative phase of the two pulses. This random choice puts an additional difficulty on top of the DPS protocol in eavesdropping, and roughly speaking the amount of the privacy amplification per sifted key bit is given by $h\left(\nu/(L-1)\right)$, where $h(x):=-x \log_2 x - (1-x) \log_2 (1-x)$ is the binary entropy function and $\nu$ the total photon number contained in $L$ pulses. This means in particular that as $L$ increases, the amount of the privacy amplification decreases. Since the privacy amplification factor does not depend on the parameters related to signal disturbance, the RRDPS protocol is in principle highly robust against channel disturbance for large $L$, and it paves a new route toward QKD over long distances and under much more harsh environment. 

The receiver's setup in the original proposal \cite{rrdps} assumed a single interferometer with an active variable delay, which is not so easy to realize with high speed and good stability. Here, as the first demonstration, we have realized a variable delay line up to 4-pulse delay by passively choosing one interferometer out of four interferometers. 




\begin{figure*}[thb]
\centerline{\includegraphics[width=\linewidth]{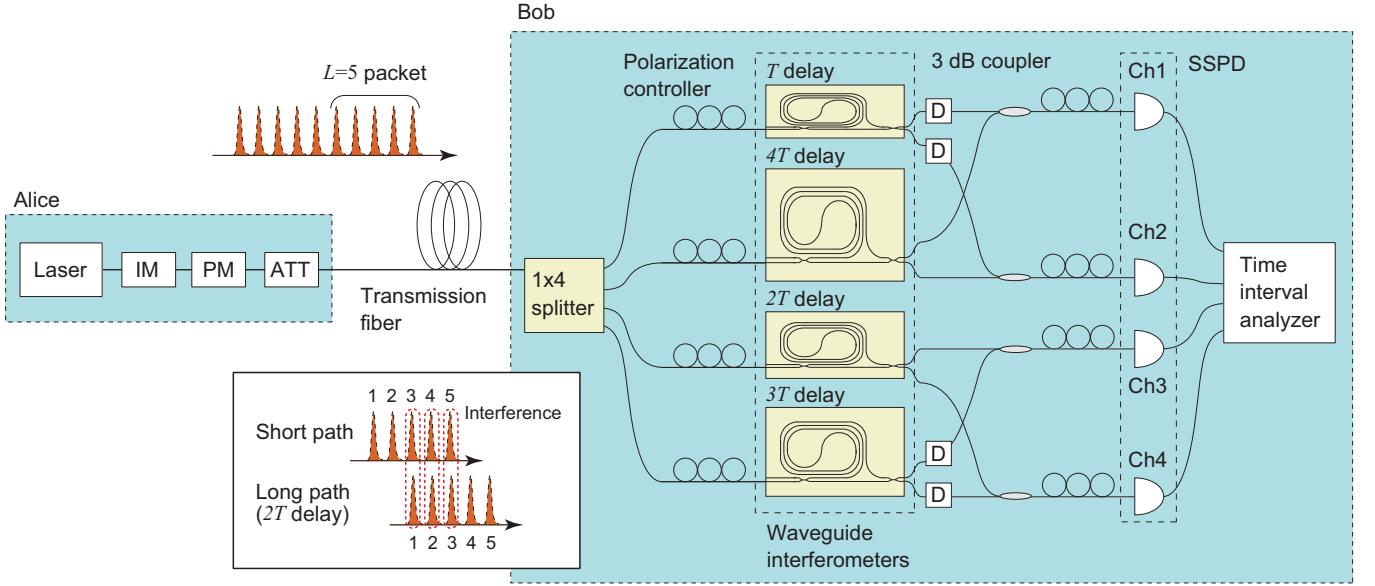}}
\caption{Experimental setup for RRDPS QKD. IM: intensity modulator, PM: phase modulator, ATT: attenuator, D: 250-ps delay line, SSPD: superconducting single photon detectors. Inset: pulse interference at $MT$-delay interferometer ($L=5$, $M=2$, $T=500$ ps). }
\label{1}
\end{figure*}

In the RRDPS protocol using a laser source, Alice prepares a packet of $L$ coherent optical pulses and modulates each pulse with a phase shift of either 0 or $\pi$. The state of the whole packet is given by
\begin{equation}
|\Psi\rangle = \bigotimes_{k=1}^L |\alpha e^{i\phi_k}\rangle_k
\end{equation}
where $|\alpha e^{i \phi_k}\rangle_k$ denotes the coherent state of the $k$th pulse. Here, the amplitude $\alpha$ is related to the average photon number $\mu$ per pulse as $|\alpha|^2 = \mu$ and the phase $\phi_k =\{0,\pi\}$. Let $\nu$ be the total photon number in a packet. Then, the probability of finding more than $\nu_{\rm{th}}$ photons in a packet is given by
\begin{equation}
{\rm{Pr}}(\nu > \nu_{\rm{th}}) \le e_{\rm{src}} := 1- \sum_{\nu=0}^{\nu_{\rm{th}}} \frac{(L\mu)^{\nu} e^{-L\mu}}{\nu !}. \label{pr}
\end{equation}
Here, $\nu_{\rm{th}}$ is an integer constant chosen in the protocol. 
The packet is transmitted over a quantum channel and sent to Bob. Bob inputs the packet into a delayed interferometer whose delay time were randomly chosen from $\{T,2T,\cdots, (L-1)T\}$, where $T$ is the temporal interval between adjacent pulses. When the $MT$-delay $(M \in \{1,2, \cdots, L-1\})$ is chosen, Bob observes photon interferences in $L-M$ slots (see inset in Fig. \ref{1}). At those time slots, a photon is output from the port 0 (1) when the phase difference between the $i$th ($i \in \{1,2, \cdots, L-M\}$) and $(i+M)$th pulses was 0 ($\pi$). Bob recognizes a photon detection events as one bit of sifted key when they are in the interference slots, and records the time slot, the detected port, and the delay used in the measurement. After repeating the same procedure many times, Bob announces over an authenticated classical channel to Alice the indices of the successfully detected packets together with the recorded time slots and delays. This is enough for Alice to determine her sifted key from her record of the phase shifts applied on the transmitted pulses.

Eve's knowledge on the sifted key can be bound by the argument in \cite{rrdps}, which is summarized as follows.
We can regard each bit of Alice's sifted key as the outcome of the $Z$-basis measurement on a virtual qubit. The probability of the qubit being found in state $|0\rangle-|1\rangle$ if it were measured in the
X basis is called the ``phase error rate," and it determines a sufficient amount of privacy amplification. It can be shown that the phase error rate is bounded by $\nu_{\rm{th}}/(L-1)$ when the number of photons in a packet is no larger than $\nu_{\rm{th}}$ \cite{rrdps}. Then, the phase error rate for the use of light source satisfying Eq.(\ref{pr}) can be no larger than 
\begin{align}
 e_{\rm{ph}} &= \frac{e_{\rm{src}}}{Q}+\left(1-\frac{e_{\rm{src}}}{Q} \right)\frac{\nu_{\rm{th}}}{L-1}, \label{eph}\\
 Q &= N/N_{\rm{em}}
\end{align}
where $N$ is the sifted key length and $N_{\rm{em}}$ is the number of packets emitted from Alice.
The first term of Eq. (\ref{eph}) corresponds to the fraction of a sifted key where the number of emitted photons in a packet may have exceeded $\nu_{\rm{th}}$ and thus has to be regarded as a phase error in the worst case scenario. The net secure key length in the asymptotic limit is then given by
\begin{equation}
G=N\left[1-f h(e_{\rm{bit}}) - h(e_{\rm{ph}}) \right], \label{securekey}
\end{equation}
where $e_{\rm{bit}}$ is the bit error rate and $f$ corresponds to the parameter related to the efficiency of the employed error correcting code. Eqs. (\ref{eph}) and (\ref{securekey}) indicates that the privacy amplification factor for the RRDPS protocol depends only on the photon number characteristics of the source, the length of the packet, and the sifted key rate, while it is independent of the error rate. In this manner, the RRDPS protocol does not require the phase error estimation using the randomly sampled test bits unlike other QKD protocols, and thus possibly simplifies the system significantly.


\begin{figure}[thb]
\centerline{\includegraphics[width=\linewidth]{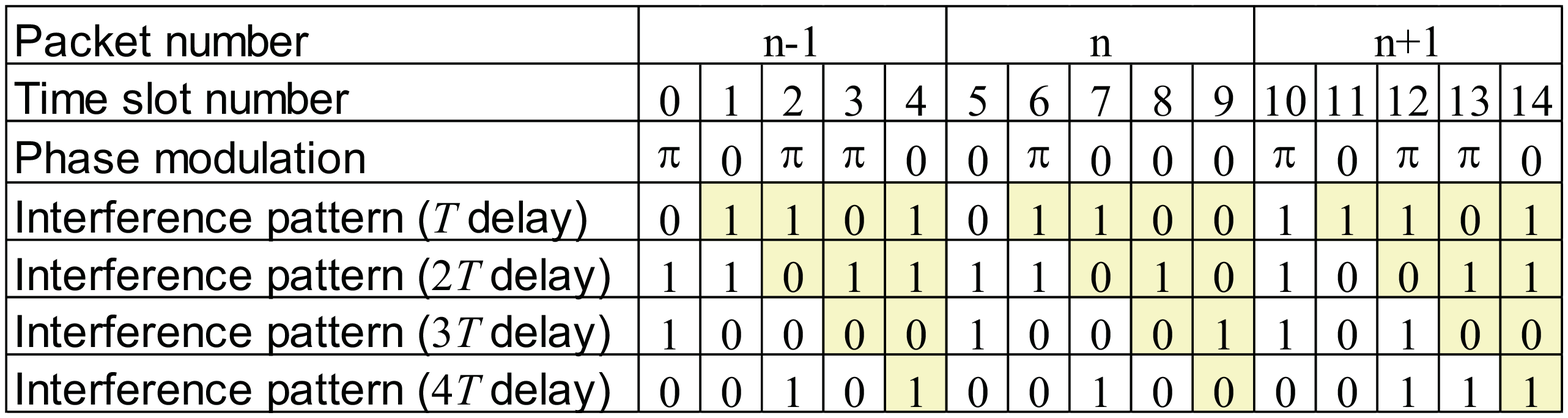}}
\caption{Example of packet configuration and interference pattern at each interferometer. }
\label{pattern}
\end{figure}

\begin{figure*}[thb]
\centerline{\includegraphics[width=\linewidth]{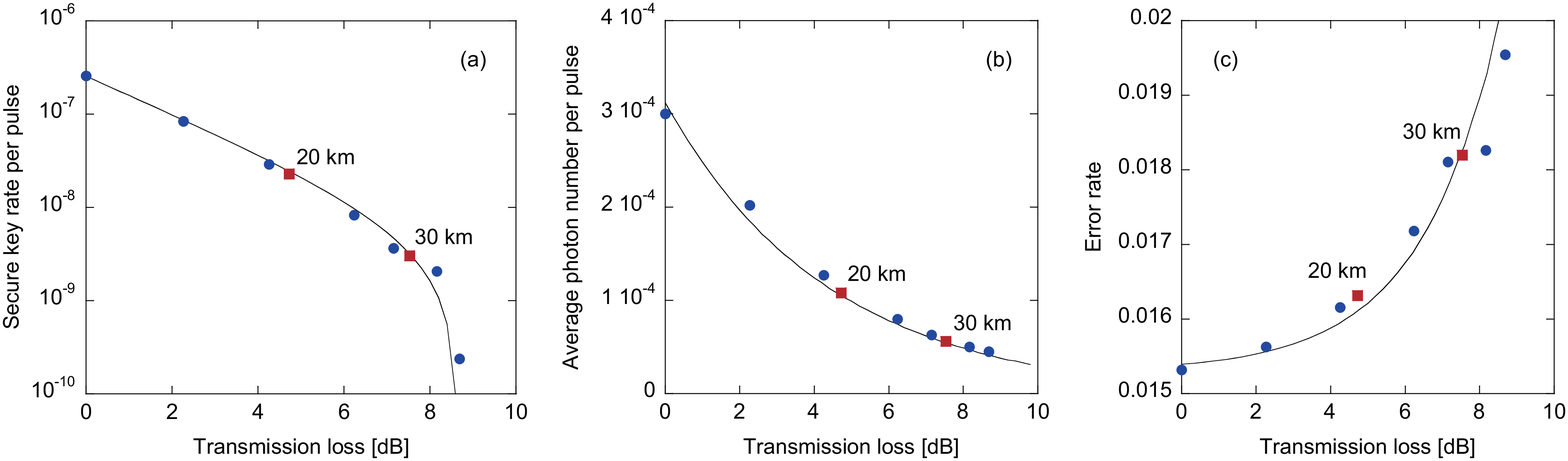}}
\caption{Experimental results. (a) Secure key rate per pulse in the limit of asymptotic key length as a function of transmission loss. (b) Average photon number per pulse used in the experiments for each transmission loss. (c) Bit error rate as a function of transmission loss. In (a) - (c), the squares and circles denote fiber transmissions and optical attenuation, respectively. The solid curves in (a)-(c) are calculated from Eq. (\ref{6}) with the average photon number per pulse optimized to maximize the secure key rate at each transmission loss (see also Method)}. 
\label{secure}
\end{figure*}

Figure \ref{1} shows the experimental setup. Alice generated optical pulses with 60 ps temporal width at a clock frequency of 2 GHz by modulating a 1551.1-nm coherent light from an external cavity diode laser using a lithium niobate intensity modulator. The time interval between the adjacent pulses is $T=0.5$ ns. 
The phases of the pulses were modulated by 0 or $\pi$ using a lithium niobate phase modulator driven by a signal from a 2-GHz pseudo-random bit pattern (PRBS) generator. Then, the pulse train was launched into an optical attenuator so that the average photon number per pulse became much less than 1. We defined a packet for every 5 pulses ($L=5$) in the pulse train, and assigned ``packet number" and ``time slot number" to each packet and pulse for the data processing (Fig. \ref{pattern}). The packets were then sent to Bob through a transmission fiber.  

Since we set $L=5$, Bob needs to randomly choose one from four delay length $\{T,2T,3T,4T\}$ in a variable-delay interferometer in the implementation in the original paper \cite{rrdps}. In this experiment, we employed a passive configuration that composed of a 1-input, 4-output optical splitter followed by 4 interferometers that have $\{0.5,1.0,1.5,2.0\}$-ns temporal delays. Note that a similar multi-interferometer configuration was proposed to increase the probability of error caused by the intercept-and-resend attack against the DPS protocol \cite{extend}.
Since a valid outcome in the RRDPS protocol is generated when a packet contains a single photon, the role of the optical splitter can be regarded as routing the packet randomly to one of the 4 interferometers. The delayed interferometers used in the experiment were fabricated using silica waveguide, and thus a precise and stable phase adjustment was possible \cite{honjo}.  
The photons from two output ports of each interferometer were then received by superconducting single photon detectors (SSPD) based on NbN nanowires (Scontel). Although 8 detectors are needed for the passive configuration with 4 interferometers, we employed a temporal multiplexing technique to reduce the required number of detectors by half. As shown in Fig. \ref{1}, a port from $T$-delay ($3T$-delay) interferometer was delayed by 250 ps and combined with that from $4T$-delay ($2T$-delay) interferometer using a 3-dB coupler, and input into an SSPD. The drawback of this temporal multiplexing is an additional 3-dB loss induced by the coupler.  The detection signals from each SSPD were input into a time interval analyzer (Picoquant) for recording the detection time instances and which-detector information. The detection efficiencies of the SSPDs were around 19\% at a dark count rate of $\sim$ 2 cps. The overall system loss (including the detection efficiencies of SSPDs) was 12.7 dB. We applied a 200-ps time window to the obtained data to reduce the contribution of dark counts.

Figure \ref{pattern} is an example of interference pattern that is observed by Bob. Although we can observe interference in every time slot (as in original DPS protocol), we only use detection events that are obtained from the interference between two pulses within the same packet, which are shown by yellow cells in Fig. \ref{pattern}. 
When we observe only one detection event in the yellow cells for a packet, we record the event as a new bit of the sifted key. 
We can estimate the frequency of Bob receiving two or more photons in a packet by observing the ``double clicks" between the detection events at detectors ``Ch 1 or Ch 2" and ``Ch 3 or Ch 4" in yellow cells in a packet. Note that such an event does not produce a bit of the sifted key. When the number of double-click events is $N_{\rm{d}}$, the fraction in the $N$-bit sifted key that originated from the multi-photon received packets is no larger than $q_{\rm{d}}:=8 N_{\rm{d}}/N$. 
The modified secure key length in the asymptotic limit that takes account of $q_{\rm{d}}$ is given by (see Method for details)
\begin{align}
G &= N[1-fh(e_{\rm{bit}}) - q_{\rm{d}} - (1-q_{\rm{d}}) h(e'_{\rm{ph}})], \label{6} \\
e'_{\rm{ph}} &= \frac{e_{\rm{src}}}{Q(1-q_{\rm{d}})} + \left(1-\frac{e_{\rm{src}}}{Q(1-q_{\rm{d}})} \right) \frac{\nu_{\rm{th}}}{L-1}. \label{7}
\end{align}
The parameter $q_{\rm{d}}$ was included in the calculation of the amount of privacy amplification in our experiments, but the effect was almost negligible since $q_{\rm{d}}$ was less than $10^{-3}$ in all the runs.


We undertook key generation experiments with various channel loss values. At each loss, we performed sifted key generation for an effective data acquisition time of 260 s. We optimized the average photon number per pulse for each loss as in Fig. \ref{secure} (b) to maximize the secure key length based on the calculation described in Method. Figure \ref{secure} (c) shows the error rate as a function of channel loss. The error rate at 0 channel loss was 1.5\%, which was limited by the imperfect extinction ratio of the delayed interferometers. The asymptotic secure key rate obtained in the experiment is plotted in Fig. \ref{secure} (a), where squares and circles denote fiber transmissions and optical attenuation, respectively. Here, we calculated secure key rates using Eq. (\ref{6}) with experimentally obtained sifted key lengths, error rates and the numbers of double-click events, with the assumption of $f=1.1$. The results indicate that asymptotically we can generate secure keys up to 8.7 dB channel loss, and secure key distribution over a 30-km fiber was realized.

\begin{figure}[thb]
\centerline{\includegraphics[width=\linewidth]{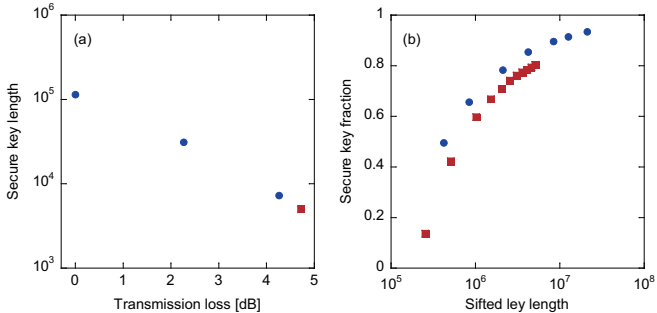}}
\caption{Results of finite-key security analysis. (a) Secure key length as a function of transmission loss for a effective data acquisition time of 260 s. Squares: 20-km fiber transmission, circles: optical attenuation. (b) Fraction of the secure key length obtained with finite-key security analysis relative to the asymptotic secure key length as a function of sifted key length. Circles and squares denote 0-dB transmission loss and 20-km fiber transmission.}
\label{finite}
\end{figure}

We then have analyzed the security accommodating the effect arising from the finiteness of the sifted key length, with a choice of the security parameter $d=2^{-50}$. The detailed procedure of finite-key security analysis is shown in Method. 
We first performed the analysis on the same data shown in Fig. \ref{secure}. The obtained secure key length is shown in Fig. \ref{finite} (a). It shows that we successfully generated secure keys up to 20 km of fiber (4.7 dB loss) with finite-key effect considered. 
We also undertook measurements with varied data acquisition times at 0-dB transmission loss and 20-km fiber transmission to assess the sifted key length dependence on the secure key. 
The result is shown in Fig. \ref{finite} (b), where the fraction of the secure key length obtained with finite-key security analysis relative to the asymptotic secure key length is plotted as a function of the sifted key length. When the fiber length is 0 (denoted by circles), the fraction was 50\% and 93\% for sifted key lengths of 420 kbits (effective data acquisition time: 26 s) and 21 Mbits (1300 s), respectively. 
We also obtained a secure key fraction of 80\% at 20-km fiber length with a sifted key length of 5.1 Mbits taken in a 2600-s data acquisition time. 


The analysis in \cite{rrdps} predicted that the performance of the RRDPS protocol can be superior to BB84 protocols with decoy states at large error rate. 
However, in our experiment with $L=5$ and the baseline system error of 1.5\%, the secure key compression factor in the asymptotic limit $1-fh(e_{\rm bit})-h(e_{\rm ph})$ is $\sim 0.94$ even if $e_{ph}$ is assumed to be 1/4 (i.e. $\nu_{th}=1$ and no multi-photons emitted in a packet). This means that our experiment was already close to the edge of secure key generation with the relatively small $L$. 
By using larger $L$, we can improve the performance of a RRPDS system significantly. 
Let us assume an RRPDS system with a delayed interferometer whose delay length can be actively changed as in the original proposal \cite{rrdps}.
We assumed that the dark count rate per detector, system loss and transmission loss are 2 cps, 12.7 dB and 4.7 dB (corresponding to 20 km fiber in our experiment), respectively, and calculated the maximum system error rate with which we can obtain secure key for each $L$. The result is shown in Table S1 in Supplementary Information. It is obvious that the tolerable error rate becomes much higher as we increase $L$. 
The finite-key effect is also improved for a large $L$. For example, at 20-km fiber length and $L=32$, 60\% of the asymptotic key rate is achieved with the sifted key length of only $3 \times 10^4$. 
Thus, if we implement larger $L$, we can expect significant improvement in the performance. 

The use of optical waveguide technologies will facilitate the implementation of a larger number of delays. Silica-waveguide-based optical delay lines whose lengths are as long as 4 m with a footprint of only 63 mm $\times$ 12 mm have already been reported \cite{4m}. With a 10-GHz clock QKD system which has been reported in \cite{np}, we can implement $\sim 200T$ delay line with a 4-m waveguide. By incorporating with a fast optical switch whose bandwidth can be more than 10 GHz using silica-lithium niobate hybrid integration technology \cite{yamazaki}, we can realize a RRDPS system with $>$100 delays.


Note that the present QKD setup is not ``secure'' as
a whole system, which is mainly because we employed
a PRBS as the phase modulation signal. In addition,
we need to implement error correction and privacy amplification
to claim the realization of full-QKD system.
The former problem may be solved by the use of quantum random number 
generators, and the latter can be solved by the direct application of the
existing software/hardware for the key distillation. 
Other issues in the present setup are security loopholes, that is, we have assumed that all the devices in the experiment operate as the theory require. In particular, we have assumed that there is no side-channel. Unfortunately, however, this is not the case in practice. For instance, the detection unit might be fragile against the so-called blight pulse illumination attacks, and imperfect phase/amplitude modulation by Alice might degrade the performance of the present setup. These are shared issues in most of the QKD experiments, and we hope in future that there will come out counter-measures to waive these problems. Even under these issues, the importance of the present experiments lies in the fact it confirmed the possibility of key generation with the fundamentally new QKD scheme, which paves a new route toward QKD systems under harsher environment.

\section*{Methods}

\noindent
{\bf Security proof outline.} 
We show an outline of a security proof against general attacks for the implementation used in the present experiment.
There are two points which differentiate this experiment from one proposed in \cite{rrdps}.
First, Bob does not use an interferometer with an actively-controlled variable delay line but employs a passive optical splitter followed by several interferometers with various temporal delays. 
As long as the number of photons received in a packet is one, this change does not make any difference in security proof because the positive operator-valued measure (POVM) characterizing this measurement is the same with that using the variable delay line.
Second, Bob cannot measure the number of photons directly because he uses threshold detectors, whereas the proof in \cite{rrdps} assumes the use of detectors that can discriminate a single photon from two or more photons. 
This is the point where extra consideration is needed, since we need to estimate the number of the received packets containing one photon using threshold detectors. 

We define a double-click event as a coincidence between a detection event at ``Ch1 or Ch2'' and one at ``Ch3 or Ch4'', both in the interference slots (yellow cells in Fig. \ref{pattern}) of the same packet.
We assume that all 4 detectors have the same detection efficiency.
This can be modeled by detectors with unit efficiency and a common linear absorber in front of Bob's apparatus.
In Table \ref{Methods-prob-MZI}, we show the probability that a photon in each pulse goes to each interferometer and is detected in the interference slots.
We see that the probability that a photon in a packet is detected at ``Ch1 or Ch2'' is 1/4 regardless of which pulse the photon is in.
The same is true for ``Ch3 or Ch4''.
This means that a lower bound of the probability of a double-click event is 1/8 when there are two or more photons received in a packet.

\begin{table}
\begin{center}
 \caption{The probability that a photon in each pulse goes to each interferometer and is detected in the interference slots.}
\label{Methods-prob-MZI}
 \begin{tabular}{|c|c|c|c|c|c|}
\hline
	&1st&2nd&3rd&4th&5th \\
\hline
	$T$   delay (``Ch1 or Ch2'')&1/8&1/4&1/4&1/4&1/8	\\ \hline
	$2T$ delay (``Ch3 or Ch4'')&1/8&1/8&1/4&1/8&1/8	\\ \hline
	$3T$ delay (``Ch3 or Ch4'')&1/8&1/8&0    &1/8&1/8	\\ \hline
	$4T$ delay (``Ch1 or Ch2'')&1/8&0    &0    &0    &1/8	\\ \hline
 \end{tabular}
\end{center}
\end{table}

Although the actual measurement device cannot discriminate the photon number, it is in principle possible to measure the total photon number in each packet in front of Bob's apparatus without affecting any procedure in the protocol.
Hence we may, in principle, tag every packet that includes two or more photons.
Let $N_{\rm{mB}}$ be the number of tagged ones among the $N_{\rm{em}}$ packets received by Bob, and $N_{\rm{d}}$ be the observed number of double-click events.
In the asymptotic limit of large $N_{\rm{mB}}$, we may assume that $N_{\rm{mB}}$ is no larger than $8N_{\rm{d}}$.
The secure key is obtained by subtracting the same number of bits \cite{gllp} in the privacy amplification, leading to a modified formula for the secure key length given by Eqs. (\ref{6}) and (\ref{7}).

\noindent
{\bf Performance estimation with experimental parameters.} 
In order to estimate the performance,
we need to assume a channel model to determine the expected asymptotic values of $Q$, $e_{\rm{bit}}$, and $p_{\rm{d}}:=N_{\rm{d}}/N_{\rm{em}}$. 
We denote $d_{\rm{c}}$ and $e_{\rm{sys}}$ as the dark count rate of the SSPD (per channel) and the baseline system error rate, respectively.
We then assume that the sifted key rate per packet and the bit error rate are approximated with the following equations. 
\begin{eqnarray}
Q &\simeq& \frac{L\mu \eta}{2} + (L-1)L d_{\rm{c}} \\
e_{\rm{bit}} &\simeq& \frac{ \frac{L\mu \eta}{2} e_{\rm{sys}} + (L-1)Ld_{\rm{c}}/2}{Q}\\
p_{\rm{d}} &\simeq& (L \mu \eta)^2/16 \label{11}
\end{eqnarray}
Using these assumptions, we optimize $\mu$ to obtain a secure key rate.

\noindent
{\bf Finite-key security analysis.} 
We will obtain a secure key length $G_{\rm{f}}$ and a security parameter $d$ from 6 protocol parameters $\bar{N}_{\rm{d}}, \bar{N}_{\rm{mB}},\bar{N}_{\rm{mA}},\bar{N}_{\rm{ph}}, s_{\rm{x}},$ and $s_{\rm{z}}$.
As the security parameter $d$, we adopt half the trace distance between the actual state $\rho_{ABE}$  and an ideal state $\rho_{ABE}^{\rm ideal}$.
We define a function for the tail distribution for finding no fewer than $k$ successes in a binomial distribution as 
\begin{equation}
 f(k;n,p) := \sum_{j\geq k} p^j (1-p)^{n-j} \binom{n}{j}.
\end{equation}
If $k/n>p$, it is bounded by $f(k;n,p) \le 2^{-nD(k/n \parallel p)}$, where  $D(q \parallel p):=q \log_2 (q/p) + (1-q) \log_2 \{(1-q)/(1-p)\}$.
The similar function for no more than $k$ successes is
\begin{equation}
 g(k;n,p) := \sum_{j\leq k} p^j (1-p)^{n-j} \binom{n}{j},
\end{equation}
which is bounded by $g(k;n,p) \le 2^{-nD(k/n \parallel p)}$ if $k/n < p$.

After a sifted-key generation session is finished, Bob first examines the number of double-click events $N_{\rm{d}}$. If it is larger than a pre-defined threshold $\bar{N}_{\rm{d}}$,
Alice and Bob discard the key.

We analyze two cases separately according to the number $N_{\rm{mB}}$ of multi-photon packets 
(the tagged packets) received by Bob's apparatus.
If $N_{\rm{mB}}$ is larger than a pre-defined threshold $\bar{N}_{\rm{mB}}$,
Alice and Bob should discard the sifted key at least with a probability $1-\epsilon_1$,
where $\epsilon_1$ is defined as
\begin{equation}
\label{methods-epsilon1}
\epsilon_1 := g(\bar{N}_{\rm{d}};\bar{N}_{\rm{mB}},1/8).
\end{equation}
Since producing no key is regarded as an ideal state, the security parameter is bounded as $ d  \leq \epsilon_1$.

For the case with $N_{\rm{mB}} \leq \bar{N}_{\rm{mB}}$, we can choose $N':=N-\bar{N}_{\rm{mB}}$ untagged packets from the $N$ packets resulting in the $N$-bit sifted key.
We then apply the security argument in \cite{rrdps} to these $N'$ packets, which is summarized as follows.
For each of the $N_{\rm{em}}$ packets emitted from Alice, we may assume that Alice is left with $L$ qubits such that measurement of each qubit 
in the standard basis $\{|0\rangle,|1\rangle\}$ would reveal the corresponding phase shift $\{0,\pi\}$.
On the other hand, the statistics of the $L$ qubits measured in the $\{|+\rangle,|-\rangle\}$ basis, where $|\pm\rangle = \frac{1}{\sqrt{2}}\left(|0\rangle \pm |1\rangle \right)$, 
are related to the property of the light source.
Let $n_-$ be the number of qubits found in state $|-\rangle$.
Then we have $\text{Prob}(n_-  > \nu_{\rm{th}}) \leq e_{\rm{src}}$.
Whenever Bob succeeds in detection, Alice may convert the $L$ qubits into
a single qubit (we call it a sifted qubit) and then measures it in the
standard basis to determine her sifted key bit.
For the purpose of proving the security, we ask what happens if Alice measures the sifted
qubit in $\{|+\rangle,|-\rangle\}$ basis instead.  We call the outcome for
state $|-\rangle$ a phase error. It was shown \cite{rrdps} that the probability of
obtaining a phase error is no larger than $n_-/(L-1)$.
Then the security of the final key generated from the $N'$ sifted key bits can be related to
the failure probability $p_{\rm{fail}}$ of correcting the phases errors in the corresponding $N'$ sifted qubits.

The total number of phase errors in the $N'$ sifted qubits is estimated as follows.
For each of the $N_{\rm{em}}$ packets emitted from Alice, $n_-$ exceeds $\nu_{\rm{th}}$ at most with probability $e_{\rm{src}}$ . Therefore, the number $N_{\rm{mA}}$ of
packets with $n_->\nu_{\rm{th}}$ is no larger than a pre-defined threshold value $\bar{N}_{\rm{mA}}$, except for a small probability defined by
\begin{equation}
\label{methods-epsilon2}
\epsilon_2 := f(\bar{N}_{\rm{mA}};N_{\rm{em}},e_{\rm{src}}).
\end{equation}
This means that, except for the probability $\epsilon_2$, at least $N'':=N'-\bar{N}_{\rm{mA}}$ packets among the $N'$ sifted packets satisfy  $n_- \leq \nu_{\rm{th}}$.
We define $N_{\rm{ph}}$  as the number of phase errors in the $N''$ packets.
The number $N_{\rm{ph}}$ should be no larger than a pre-defined threshold $\bar{N}_{\rm{ph}}$ 
with a probability no smaller than $1-\epsilon_2-\epsilon_3$, where $\epsilon_3$ is defined as 
\begin{equation}
\label{methods-epsilon3}
\epsilon_3 := f \left(\bar{N}_{\rm{ph}};N'',\frac{\nu_{\rm{th}}}{L-1} \right). 
\end{equation}
Since it makes the total number of phase errors in the $N'$ sifted qubits no larger than $\bar{N}_{\rm{mA}}+\bar{N}_{\rm{ph}}$, a virtual phase error correction with a syndrome length
\begin{equation}
 N' h\left(\frac{\bar{N}_{\rm{mA}} + \bar{N}_{\rm{ph}}}{N'}\right) + s_{\rm{x}}
\end{equation}
will succeed at least with $1-p_{\rm{fail}}$,
where $p_{\rm{fail}}:=\epsilon_2+\epsilon_3+\eta_{\rm{x}}$ and
$\eta_{\rm{x}}:=2^{-s_{\rm{x}}}$.

As a result, for the case with $N_{\rm{mB}}\leq \bar{N}_{\rm{mB}},$ the secure 
key length with the finite-length effect taken into account is given by
\begin{equation}
G_{\rm{f}} = N' - f Nh(e_{\rm{bit}})-s_{\rm{z}} -  N'h\left(\frac{\bar{N}_{\rm{mA}} + \bar{N}_{\rm{ph}}}{N'}\right)-s_{\rm{x}}, \label{gf}
\end{equation}
where $s_{\rm{z}}=-\log_2\eta_{\rm{z}}$ bits are used for a verification of the bit error correction.
The security parameter is bounded as $d \leq \eta_{\rm{z}} + \sqrt{2} \sqrt{p_{\rm{fail}}}$ \cite{rmp2}.

Since any attack by Eve can be regarded as a mixture of the two cases analyzed above, $d$ is always bounded as
\begin{equation}
\label{methods-d-bound}
 d \leq \max(\epsilon_1, \eta_{\rm{z}} + \sqrt{2} \sqrt{\epsilon_2 + \epsilon_3 + \eta_{\rm{x}}}).
\end{equation}

To obtain the numerical results in this paper, we set $\bar{N}_{\rm{d}}$ to be $N_{\rm{em}} p_{\rm{d}} + 3 \sqrt{N_{\rm{em}} p_{\rm{d}}}$,
which makes the probability of discarding the protocol negligibly small.
We fixed $\epsilon_1, \epsilon_2, \epsilon_3, \eta_{\rm{x}},$ and $\eta_{\rm{z}}$ to be $2^{-50}, 2^{-103}/3,2^{-103}/3,2^{-103}/3$, and $2^{-51},$ respectively.
Then we determined the values of $\bar{N}_{\rm{mB}},\bar{N}_{\rm{mA}},\bar{N}_{\rm{ph}}, s_{\rm{x}},$ and $s_{\rm{z}}$ to satisfy Eqs. (\ref{methods-epsilon1}), (\ref{methods-epsilon2}),  (\ref{methods-epsilon3}),
and the definitions of $\eta_{\rm{x}}$ and $\eta_{\rm{z}}$.
This achieves the condition $d\leq 2^{-50}$.

\section*{Acknowledgements}
The authors thank Yoshihisa Yamamoto for fruitful discussions. 
This work was funded in part by ImPACT Program of Council for Science,
Technology and Innovation (Cabinet Office, Government of Japan), Photon Frontier Network Program (MEXT), and the National Institute of Information and Communications Technology (NICT) of Japan.

\section*{Author contributions}

H.T. designed and constructed the experimental setup, and performed the QKD experiments. 
T.S. and M.K. designed the detailed procedure for the secure key generation. 
H.T., T.S. and K.T. undertook data analysis. 
M.K. led the whole project. 
All authors discussed the result and wrote the paper.

\section*{Corresponding authors}

Hiroki Takesue (takesue.hiroki@lab.ntt.co.jp) and Toshihiko Sasaki (sasaki@qi.t.u-tokyo.ac.jp ).


\clearpage

\section*{Supplementary Information}

\subsection*{S1: Performance improvement with larger number of delays} \label{appd}
We assume a setup that uses a single interferometer with an actively controlled variable delay followed by two single photon detectors as proposed in \cite{rrdps}. 
With this setup, the sifted key rate per packet and the bit error rate are given by 
\begin{eqnarray}
Q &=& \frac{L\mu \eta}{2} + L d_{\rm{c}}, \label{qsi}\\
e_{\rm{bit}} &=& \frac{ \frac{L\mu \eta}{2} e_{\rm{sys}} + Ld_{\rm{c}}/2}{Q}. \label{esi}
\end{eqnarray}
We assume that the system loss, the fiber loss (20 km), and the dark count rate per detector are the same as the experiment, 12.7 dB, 4.7 dB, and 2 cps, respectively.
We also assume the use of 200-ps time window.
Using Eqs. (\ref{6})-(\ref{7}), (\ref{11}), (\ref{qsi}) and (\ref{esi}), we calculated the maximum system error rate for each $L$ with which we can obtain any key.
The result is shown in Table S1.

\begin{table}[b]
TABLE S1: The maximum system error rate to obtain any secure key for each $L$

\begin{tabular}{c|c}
\hline
$L$ & $ e_{\rm{sys}}$\\
\hline
5 & $2.3\%$\\
16 & $13.3\%$ \\
32& $18.6\%$\\
64& $22.1\%$\\
128& $24.4\%$\\ 
\hline
\end{tabular} 
\end{table}


\begin{thebibliography}{99}

\bibitem{bb84} Bennett, C. H. \& Brassard, G. in {\it Proc. IEEE Int. Conf. on Computers, Systems and Signal Processing} 175-179 (IEEE Press, 1984).

\bibitem{e91} Ekert, A. K. Quantum cryptography based on Bell's theorem. {\it Phys. Rev. Lett.} {\bf 67}, 661-663 (1991).

\bibitem{b92} Bennett, C. H. Quantum cryptography using any two nonorthogonal states. {\it Phys. Rev. Lett.} {\bf 68}, 3121-3124 (1992).

\bibitem{huttner} Huttner, B., Imoto, N., Gisin, N., \& Mor, T. Quantum cryptography with coherent states. {\it Phys. Rev. A} {\bf 51}, 1863-1869 (1995). 

\bibitem{cv} Jouguet, P., Kunz-Jacques, S.,  Leverrier, A., Grangier, P., 
\& Diamanti, E., Experimental demonstration of long-distance continuous-variable quantum key distribution. {\it Nature Photon.} {\bf 14}, 378-381 (2013).  

\bibitem{kyo} Inoue, K., Waks, E., \& Yamamoto, Y. Differential phase shift quantum key distribution. {\it Phys. Rev. Lett.} {\bf 89}, 037902 (2002). 

\bibitem{sarg} Scarani, V., Acin, A., Ribordy, G.\& Gisin, N. Quantumcryptography protocols robust against photon number splitting attacks for weak laser pulse implementations. {\it Phys. Rev. Lett.} {\bf 92}, 057901 (2004).

\bibitem{cow} Stucki, D., Brunner, N., Gisin, N., Scarani, V. \& Zbinden, H. Fast and simple one-way quantum key distribution. {\it Appl. Phys. Lett.} {\bf 87}, 194108 (2005).

\bibitem{secoqc} Peev, M. et al. The SECOQC quantum key distribution network in Vienna. {\it New J. Phys.} {\bf 11}, 075001 (2009).

\bibitem{tokyoqkd} Sasaki, M. et al. Field test of quantum key distribution in the Tokyo QKD network. {\it Opt. Express} {\bf 19}, 10387 (2011). 

\bibitem{rmp} Gisin, N., Ribordy, G., Tittel, W., \& Zbinden, H. Quantum cryptography.  {\it Rev. Mod. Phys.} {\bf 74}, 145-195 (2002). 

\bibitem{rmp2} Scarani, V., Bechmann-Pasquinucci, H., Cerf, H. J., Dusek, M., Lutkenhaus, N., \& Peev, M. The security of practical quantum key distribution. {\it Rev. Mod. Phys.} {\bf 81}, 1301-1350 (2009). 

\bibitem{rrdps} Sasaki, T., Yamamoto, Y. \& Koashi, M. Practical quantum key distribution protocol without monitoring signal disturbance. {\it Nature} {\bf 509}, 475-478 (2014). 

\bibitem{extend} Honjo, T. \& Inoue, K. Differential-phase-shift quantum key distribution with an extended degree of freedom. {\it Opt. Lett.} {\bf 31}, 522-524 (2006). 

\bibitem{honjo}  Honjo, T., Inoue, K., \& Takahashi, H. Differential-phase-shift quantum key distribution experiment with a planar light-wave circuit Mach-Zehnder interferometer. {\it Opt. Lett.} {\bf 29}, 2797-2799 (2004).


\bibitem{4m} Kwack, M. J., Oyama, T., Hashizume, Y., Mino, S., Zaitsu M., Tanemura, T., Nakano, Y. Compact optical buffer module for intra-packet synchronization based on InP 1x8 switch and silica-based delay line circuit. {\it IEICE Trans. Electron.} {\bf E96-C}, 738 (2013). 

\bibitem{np} Takesue, H., Nam, S. W., Zhang, Q., Hadfield, R. H., Honjo, T., Tamaki K., \& Yamamoto, Y. Quantum key distribution over 40 dB channel loss using superconducting single-photon detectors.  {\it Nature Photon.} {\bf 1}, 343 (2007).

\bibitem{yamazaki} Yamazaki, H., Yamada, T., Goh, T., Sakamaki, Y., \& Kaneko, A. 64QAM modulator with a hybrid configuration of silica PLCs and LiNbO$_3$ phase modulators. {\it IEEE Photon. Technol. Lett.} {\bf 22}, 344 (2010). 

\bibitem{gllp}
 Gottesman, D., Lo, H.-K.,
  L\"utkenhaus, N. \& Preskill, J.
 Security of quantum key distribution with imperfect device.
 Quant. Inf. Comp.
  \textbf{4}, 325 (2004).

\end{thebibliography}
\end{document}